\documentclass[aps,floats,showpacs,nofootinbib,floatfix]{revtex4}%
\usepackage{amsfonts}
\usepackage{amsmath}
\usepackage{graphicx}
\usepackage{amssymb}%
{\catcode`\|=\active \gdef\Braket#1{\left<\mathcode`\|"8000\let|\bravert
{#1}\right>}}
\def\bravert{\egroup\,\vrule\,\bgroup}

\begin{document}
\title{Pion Parton Distributions in a non local Lagrangian.}
\author{S. Noguera}
\email{Santiago.Noguera@uv.es}
\author{V. Vento}
\email{Vicente.Vento@uv.es}
\affiliation{Departamento de Fisica Teorica and Instituto de F\'{\i}sica Corpuscular,
Universidad de Valencia-CSIC, E-46100 Burjassot (Valencia), Spain.}
\date{\today }

\begin{abstract}

\end{abstract}
\begin{abstract}
We use phenomenological nonlocal Lagrangians, which lead to non trivial forms
for the quark propagator, to describe the pion. We define a procedure based on
previous studies on non local lagrangians for the calculation of the pion
parton distributions at low $Q^{2}$ . The obtained parton distributions
fulfill all the wishful properties. Using a convolution approach we
incorporate the composite character of the constituent quarks. We evolve,
using the Renormalization Group, the calculated parton distributions to the
experimental scale and compare favorably with the data and draw conclusions.

\end{abstract}

\maketitle

\section{Introduction}

Parton Distributions Functions (PDFs) and Generalized Parton Distributions
(GPDs) \cite{MullerRobaschikGeyerDittesHorejsiFortsch94-98, Radyushkin97,
Ji97}, relating PDFs and electromagnetic form factors, encode unique
information on the non perturbative hadron structure (for a recent review, see
\cite{Diehl03}). Any realistic model of hadron structure should be able to
calculate them. We shall proceed in here to study the parton distribution
functions of the pion in a previous developed formalism \cite{Noguera05}.

From the experimental point of view the PDFs and GPDs of the pion are
difficult to determine. Pions decay into muons or photons and therefore the
pion distribution functions can not be obtained from direct DIS experiments.
The pion parton distribution has been inferred from the Drell-Yang process
\cite{BadierNA3,BetevNA10,Conway89} and direct photon production
\cite{Aurencheetall89} in pion-nucleon and pion-nucleus collisions. These set
of experiments have been analyzed in ref. \cite{SuttonMartinRobertsStirling92}
obtaining that the fraction of moment of each valence (anti)quark in the pion
is about 0.23 for $Q^{2}=4\operatorname{GeV}^{2}.$

The parton distribution functions of the pion has been a subject of much
discussion in the literature. In \cite{BestSchafer97,
DetmoldMelnitchoukThomas03} calculations using quenched lattice QCD, for their
lowest moments, were performed. The lack of sea quarks in the approximation
implies a greater presence of valence quarks. Its PDFs
\cite{DavidsonRuizArriola, RuizArriolaBroniowski02} and GPDs
\cite{NogueraTheusslVento04} have been also calculated in the Nambu-Jona
Lasinio (NJL) model \cite{Nambu}. In the chiral limit, its quark valence
distribution is as simple as $q\left(  x\right)  =\theta\left(  x\right)
\theta\left(  1-x\right)  .$ Once evolution is taken into account, good
agreement is reached between the calculated PDFs and the experimental
results\cite{DavidsonRuizArriola}. The pion PDFs and GPDs have also been
calculated in a model with the simplest pseudoscalar coupling between the pion
and the constituent quark fields \cite{Lansberg}, and in the instanton model,
\cite{PolyakovWeiss99,Polyakov99}. In ref. \cite{PolyakovWeiss99} the chiral
limit result of the NJL model for PDFs changes such that $q\left(  x\right)  $
goes to zero for $x\rightarrow0$ and $x\rightarrow1.$ The pion PDFs have also
been calculated in a spectral quark model\cite{RuizArriolaBroniowski03}.

Also, the pion PDFs have been calculated in a statistical model, without any
dynamical assumption, obtaining quite reasonable results \cite{AlbergHenley05}
. In \cite{AnikinBinosiMedranoNogueraVento02} the GPDs of the pion are
calculated in the bag model. A relevant contribution to the calculation of the
pion GPDs on the light-front has been given by Tiburzi and Miller
\cite{TiburziMiller}, and some remarks on the use of the light-front for
calculating GPDs can be found in \cite{Simula}. Looking for a more fundamental
approach, not related to a unique model, the pion PDF has been also calculated
in the framework of Dyson-Schwinger Equations \cite{HechtRobertsSchmidt01}.

In all the previous calculations the pion was built of two valence constituent
(anti)quarks. However, these constituent (anti)quarks are themselves built of
elementary partons. The way to connect a description in terms of constituent
quarks with the description in terms of the elementary partons was described
in \cite{AltarelliCabibboMaianiPetronzio74,NogueraScopettaVento}. Hereafter,
we will call this procedure ACMP. In
\cite{AltarelliCabibboMaianiPetronzio74,Scopetta1,Scopetta2} the procedure was
applied to the nucleon case and in \cite{AltarelliPetrarcaRapuano96} it was
applied to the pion case. The same description of the constituent quarks in
terms of partons is used for the proton and for the pion.

The aim of this paper is to study the pion PDFs following the ideas developed
in ref \cite{Noguera05, NogueraTheusslVento04}. Our main objective is to
develop a calculational method for the PDFs preserving all the fundamental
symmetries, i.e. momentum conservation and gauge symmetry. Another motivation
is to understand the simple result obtained for the pion PDFs in the NJL model
in connection with our more elaborate description.

Our study is done in a lagrangian model \cite{Noguera05} built with the
following properties: (i) manifest chiral invariance for the strong part of
the interaction; (ii) a momentum dependence for the quark propagator provided
by QCD based calculations; (iii) the implementation of local gauge invariance
provides the electromagnetic interaction. This procedure allows us to have
simultaneously a coherent description of the dressed quark propagator, the
pion state, the dressed quark photon vertex and to preserve gauge symmetry. In
this sense our aim is not to fit the best possible description of the pion
PDFs, but to learn about the influence of the momentum dependence of the quark
propagator on the definition of the PDFs. The differences between our approach
and the previous calculation \cite{HechtRobertsSchmidt01} will become obvious.
We achieve our objective in a three step process. The first is to construct
the PDFs of the pion in terms of its constituents quarks, for that we will use
the model developed in ref. \cite{Noguera05}. For this model the pion form
factor was calculated obtaining good agreement with data, and the operator
form for the mesonic parton distribution was found in terms of the quark
propagator and the kernel of the BS equation. The second step is to include
the partonic structure of the constituent quarks. To do so we will use the
ACMP procedure assuming for the quarks the same structure functions as was
obtained in the analysis for the proton. The last step is to evolve (to next
to leading order), by means of the renormalization group of QCD, the resulting
parton distribution up to the value of $Q^{2}$ for which data exist
\cite{Traini}.

This paper is organized as follows. In section II we discuss the definition of
the PDFs in our formalism and give provide some technical details of the
calculation. In section III we perform an analysis of the PDFs, with models
defined from fundamental approaches, achieving a good description of the data.
In section IV we calculate the moments of the parton distribution, using three
different procedures, to check the validity of our method for calculating
parton distributions, and in section V we provide some conclusions.

\section{The constituent parton distribution functions}

In our scheme the pion arises as a solution of the two body BS and therefore
is a composite system of two \textquotedblleft constituent" valence
(anti)quarks which determine the PDFs. To calculate them the corresponding
operator was introduced in ref \cite{Noguera05} (we refer the reader to this
paper for details).

The parton distribution has three contributions:
\begin{equation}
q_{u}\left(  x\right)  =q_{u}^{\left(  1\right)  }\left(  x\right)  +\tilde
{q}_{u}^{\left(  1\right)  }\left(  x\right)  +q_{u}^{\left(  2\right)
}\left(  x\right)  . \label{02.03}%
\end{equation}

The first one, $q_{u}^{\left(  1\right)  }\left(  x\right)  $, corresponds to
the one body term
\begin{gather}
q_{u}^{\left(  1\right)  }\left(  x\right)  =-\int\frac{d^{4}p}{\left(
2\pi\right)  ^{4}}\text{ }\delta\left(  x-\frac{n}{2}\cdot\left(  2p+P\right)
\right)  \,\mathbb{T}\mathrm{r}\left[  i\,S\left(  p-\frac{1}{2}P\right)
\bar{\Gamma}^{M}\left(  p,P\right)  i\,S\left(  p+\frac{1}{2}P\right)  \right.
\nonumber\\
\left.  \Gamma_{\mu}\left(  p+\frac{1}{2}P,p+\frac{1}{2}P\right)  n^{\mu
}\,i\,S\left(  p+\frac{1}{2}P\right)  \Gamma^{M}\left(  p,P\right)  \right]
\label{02.04}%
\end{gather}
where $\mathbb{T}$r represent the trace in Dirac, color and flavor indices,
$\Gamma^{\mu}\left(  p,p^{\prime}\right)  $ is the quark-photon vertex which
in general can have a complicated structure (see \cite{Noguera05, BallChiu80,
CurtisPennington90}). Here $\Gamma^{M}\left(  p,P\right)  $ is the BS
amplitude for the meson,%

\begin{equation}
\Gamma_{\gamma\alpha}^{M}\left(  p,P\right)  =-2i\int\frac{d^{4}p^{\prime}
}{\left(  2\pi\right)  ^{4}}G_{\alpha\beta\gamma\delta}\left(  p,p^{\prime
},P\right)  \left(  i\,S\left(  p^{\prime}+\frac{1}{2}P\right)  ~\Gamma
^{M}\left(  p^{\prime},P\right)  \,i\,S\left(  p^{\prime}-\frac{1}{2}P\right)
\right)  _{\beta\delta} \label{02.01}%
\end{equation}
with $\alpha, \beta, \gamma$ and $\delta$ including spinor, color and flavor
indices. The normalization condition for the BS amplitude is:
\begin{gather}
\hspace*{-13cm}2iP^{\mu}=\nonumber\\
\int\frac{d^{4}p}{\left(  2\pi\right)  ^{4}}\mathbb{T}\text{r}\left[
\bar{\Gamma}^{M}\left(  p,P\right)  i\,\frac{\partial S\left(  p+\frac{1}
{2}P\right)  }{\partial P_{\mu}}\,\Gamma^{M}\left(  p,P\right)  \,i\,S\left(
p-\frac{1}{2}P\right)  +\bar{\Gamma}^{M}\left(  p,P\right)  i\,S\left(
p+\frac{1}{2}P\right)  \,\Gamma^{M}\left(  p,P\right)  \,i\,\frac{\partial
S\left(  p-\frac{1}{2}P\right)  }{\partial P_{\mu}}\right]  -\nonumber\\
\hspace*{-0.5cm}2i\int\frac{d^{4}p}{\left(  2\pi\right)  ^{4}}\frac
{d^{4}p^{\prime}}{\left(  2\pi\right)  ^{4}}\left[  i\,S\left(  p-\frac{1}
{2}P\right)  \bar{\Gamma}^{M}\left(  p,P\right)  i\,S\left(  p+\frac{1}
{2}P\right)  \right]  _{\alpha\gamma}\frac{\partial G_{\alpha\beta\gamma
\delta}\left(  p,p^{\prime},P\right)  }{\partial P_{\mu}}\left[  i\,S\left(
p^{\prime}+\frac{1}{2}P\right)  \,\Gamma^{M}\left(  p^{\prime},P\right)
\,i\,S\left(  p^{\prime}-\frac{1}{2}P\right)  \right]  _{\beta\delta}.
\label{02.02}%
\end{gather}

\begin{figure}[ptb]
\begin{center}
\includegraphics[
height=3.3916cm, width=8.613cm ] {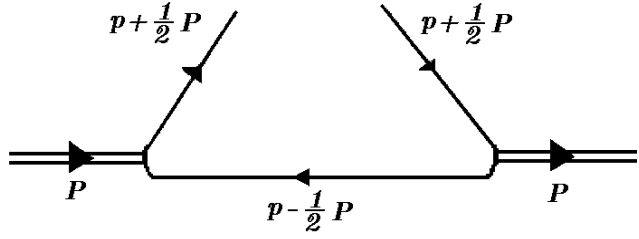}
\end{center}
\caption{Diagrammatic expression of the first one body term in Eq.(\ref{02.03}%
)}%
\label{FigPionPD1}%
\end{figure}

To calculate the PDF we need the elastic vertex $\Gamma^{\mu}\left(
p,p\right)  $\ directly related to the quark propagator by using the Ward
Identity:
\begin{equation}
\Gamma^{\mu}\left(  p,p\right)  =\frac{\partial S^{-1}\left(  p\right)
}{\partial p_{\mu}}~. \label{02.04aWardIdentity}%
\end{equation}
The pion momentum can be expressed in terms of the light-front vectors as
$P^{\mu}=\bar{p}^{\mu}+m_{\pi}^{2}n^{\mu}/2$ with $\bar{p}^{\mu}=\bar
{P}\left(  1,0,0,1\right)  /\sqrt{2},$\ $n^{\mu}=\left(  1,0,0,-1\right)
/(\sqrt{2}\bar{P}),$\ where $\bar{P}$ has a particular value for each frame
(for instance $\bar{P}=M/\sqrt{2}$ for the rest frame). This first
contribution corresponds to diagram shown in Fig.~\ref{FigPionPD1}.

\begin{figure}[ptbptb]
\begin{center}
\includegraphics[
height=2.8293cm, width=16.4132cm ] {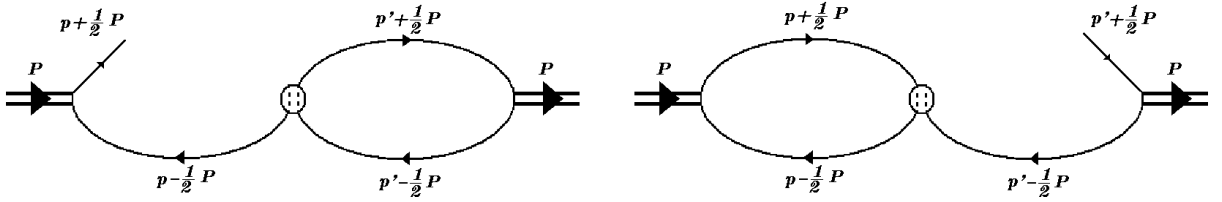}
\end{center}
\caption{Diagrammatic expression of the second one body term and the two body
term in Eq.(\ref{02.03})}%
\label{FigPionPD2}%
\end{figure}

The second term is also a one body term:
\begin{gather}
\tilde{q}_{u}^{\left(  1\right)  }\left(  x\right)  =\int\frac{d^{4}p}{\left(
2\pi\right)  ^{4}}\frac{1}{4}\delta\left(  x-\frac{n}{2}\cdot\left(
2p+P\right)  \right)  \mathbb{T}\mathrm{r}\left[  in_{\mu}~\frac{d\bar{\Gamma
}^{M}\left(  p,P\right)  }{dp_{\mu}}\,i\,S\left(  p+\frac{1}{2}P\right)
\Gamma^{M}\left(  p,P\right)  i\,S\left(  p-\frac{1}{2}P\right)  \right.
\nonumber\\
+\left.  i\,S\left(  p-\frac{1}{2}P\right)  \bar{\Gamma}^{M}\left(
p,P\right)  i\,S\left(  p+\frac{1}{2}P\right)  in_{\mu}~\frac{d\Gamma
^{M}\left(  p,P\right)  }{dp_{\mu}}\right]  \label{02.05}%
\end{gather}
while the third term is a genuine two body term given by:
\begin{gather}
q_{u}^{\left(  2\right)  }\left(  x\right)  =\int\frac{d^{4}p}{\left(
2\pi\right)  ^{4}}\int\frac{d^{4}p^{\prime}}{\left(  2\pi\right)  ^{4}
}\,\left[  i\,S\left(  p^{\prime}-\frac{1}{2}P\right)  \bar{\Gamma}^{M}\left(
p^{\prime},P\right)  i\,S\left(  p^{\prime}+\frac{1}{2}P\right)  \right]
_{\alpha\gamma}\nonumber\\
\frac{1}{2}n_{\mu}~\left\{  \left[  \delta\left(  x-\frac{n}{2}\cdot\left(
2p+P\right)  \right)  \left(  \frac{d}{dp_{\mu}^{\prime}}+2\frac{d}{dP_{\mu}
}\right)  +\delta\left(  x-\frac{n}{2}\cdot\left(  2p^{\prime}+P\right)
\right)  \left(  \frac{d}{dp_{\mu}}+2\frac{d}{dP_{\mu}}\right)  \right]
G_{\alpha\beta\gamma\delta}\left(  p^{\prime},p,P\right)  \right\} \nonumber\\
\left[  \,i\,S\left(  p+\frac{1}{2}P\right)  \Gamma^{M}\left(  p,P\right)
i\,S\left(  p-\frac{1}{2}P\right)  \right]  _{\beta\delta} \label{02.06}%
\end{gather}
These two last contributions corresponds to the diagram of
Fig~\ref{FigPionPD2} and they result from the non local character of the
interactions . In the contribution $\tilde{q}_{u}^{\left(  1\right)  }\left(
x\right)  $ the bubble integral has been performed using the BS equation,
whereas this is not possible in the contribution $q_{u}^{\left(  2\right)
}\left(  x\right)  .$



In ref. \cite{Noguera05} a particular model was developed defined by means of
a chirally invariant lagrangian:
\begin{equation}
\mathcal{L}\left(  x\right)  =\bar{\psi}\left(  x\right)  \left(
i\ \rlap{$/$}\partial-m_{0}\right)  \psi\left(  x\right)  +g_{0}\left[
J_{S}^{\dagger}\left(  x\right)  J_{S}\left(  x\right)  +\vec{J}
_{5}^{\;\dagger}\left(  x\right)  \vec{J}_{5}\left(  x\right)  \right]
+g_{p}\ J_{p}^{\dagger}\left(  x\right)  J_{p}\left(  x\right)
\label{02.07Lagrangiano}%
\end{equation}
built with the non local currents defined by
\begin{align}
J_{S}\left(  x\right)   &  =\int d^{4}y\,G_{0}\left(  y\right)  \bar{\psi
}\left(  x+\frac{1}{2}y\right)  \psi\left(  x-\frac{1}{2}y\right) \nonumber\\
\vec{J}_{5}\left(  x\right)   &  =\int d^{4}y\,G_{0}\left(  y\right)
\bar{\psi}\left(  x+\frac{1}{2}y\right)  i\vec{\tau}\gamma_{5}\psi\left(
x-\frac{1}{2}y\right) \label{02.08Currents}\\
J_{p}\left(  x\right)   &  =\int d^{4}y\,G_{p}\left(  y\right)  \bar{\psi
}\left(  x+\frac{1}{2}y\right)  \frac{1}{2}i\overleftrightarrow
{\rlap{$/$}\partial}\psi\left(  x-\frac{1}{2}y\right) \nonumber
\end{align}
where $u\overleftrightarrow{\partial}v=u\left(  \partial v\right)  -\left(
\partial u\right)  v.$ The first and second currents require the same
$G_{0}\left(  y\right)  $, in order to implement chiral invariance, while the
third current is self-invariant under chiral transformations. The scalar
current, $J_{S},$ generates a momentum dependent mass, and the last current,
the \textquotedblleft momentum" current, $J_{p},$ is responsible for the
momentum dependence of the wave function renormalization. The pseudo-scalar
current, $\vec{J}_{5},$ generates the pion pole. For simplicity, we assume
that all the $G\left(  y\right)  $ functions are real.

Let us define
\begin{equation}
G\left(  p\right)  =\int d^{4}y\,e^{iyp}G\left(  y\right)  \label{02.09}%
\end{equation}
with the normalization condition $G\left(  p=0\right)  =1.$ For simplicity we
consider that our effective theory arises from the large $N_{c}$\ limit of
$QCD$. This is equivalent to working in the Hartree approximation, i.e. only
direct diagrams will be considered.

The model introduces, in a natural way, momentum dependence to the quark mass
and to the wave function renormalization of the quark propagator:
\begin{equation}
S\left(  p\right)  =Z\left(  p\right)  \frac{\rlap{$/$}p+m\left(  p\right)  }
{p^{2}-m^{2}\left(  p\right)  +i\epsilon} \label{02.11Propagador}%
\end{equation}
where the explicit expressions for $m\left(  p\right)  $ and $Z\left(
p\right)  $ are given in the Appendix \ref{AppendixA}. The scalar current can
give a mass for the quark even if the Lagrangian contains no explicit mass
term, $m_{0}=0,$ leading to a spontaneous symmetry breaking mechanism similar
to that in the Nambu-Jona Lasinio model. On the other hand, the momentum
current gives rise to a momentum dependent wave function normalization.
However, even though the latter contributes to the mass, it is not able by
itself to break chiral symmetry spontaneously.

The model defined in Eq. (\ref{02.07Lagrangiano}) gives the following
interaction terms,
\begin{equation}
G_{\alpha\beta\gamma\delta}\left(  p^{\prime},p,P\right)  =2\ g_{0}
\ G_{0}\left(  p^{\prime}\right)  G_{0}\left(  p\right)  \left[
\delta_{\gamma\alpha} \delta_{\delta\beta}+ \left(  i\vec{\tau}\gamma
_{5}\right)  _{\gamma\alpha}\left(  i\vec{\tau}\gamma_{5}\right)
_{\delta\beta}\right]  +2\ g_{p}\ G_{p}\left(  p^{\prime}\right)  G_{p}\left(
p\right)  \left(  \rlap{$/$}p^{\prime}\right)  _{\gamma\alpha} \left(
\rlap{$/$}p\right)  _{\delta\beta} \label{02.19}%
\end{equation}
and the corresponding BS equation (\ref{02.01}) can be easily solved leading
to:
\begin{equation}
\Gamma^{M=\pi^{i}}\left(  p,P\right)  =\,i\,\gamma_{5}\,\tau^{i}\,g_{\pi
qq}G_{0}\left(  p\right)  \label{02.15PionBSA}%
\end{equation}
In Appendix \ref{AppendixA} we give the explicit expressions fixing the pion
mass and the normalization constant $g_{\pi qq}$. This model realizes the
Goldstone theorem and therefore $m_{\pi}$ goes to zero when $m_{0}$ vanishes.
This is a consequence of the use of the same $G_{0}\left(  p\right)  $
function in the scalar current, $J_{S},$ and pseudoscalar current, $\vec
{J}_{5},$ in Eq. (\ref{02.08Currents}). On the other hand, we have no
constrain on $G_{p}\left(  p\right)  $ imposed by chiral symmetry.

Inserting Eqs.(\ref{02.19}) and (\ref{02.15PionBSA}) in Eq. (\ref{02.06}) we
obtain
\begin{equation}
q_{u}^{\left(  2\right)  }\left(  x\right)  =0. \label{02.20}%
\end{equation}
Summarizing, our scheme leads to two contributions: the standard one,
associated with the triangle diagram, and defined in (\ref{02.04}) and a new
contribution defined in Eq. (\ref{02.05}). This latter contribution arises
from the non locality of the currents in our model, it is also a triangle
diagram, from the point of view of QCD, but once written in terms of the BS
amplitudes it gets a different structure.

We now proceed to the explicit calculation of the PDF. First of all we can use
the $\delta$-function present in Eqs. (\ref{02.04}) and (\ref{02.05}) in order
to integrate over $p^{3}:$
\begin{equation}
\delta\left(  x-\frac{n}{2}\cdot\left(  2p+P\right)  \right)  =\sqrt{2}\bar
{P}~\delta\left(  \sqrt{2}\bar{P}\left(  x-\frac{1}{2}\right)  -p^{0}
-p^{3}\right)
\end{equation}
With this integration, all the quantities appearing in Eqs. (\ref{02.04}) and
(\ref{02.05}) become linear in $p^{0}.$ In order to show explicitly that the
results do not depend on the choice of $\bar{P}$ we use the following
variable
\begin{equation}
2\sqrt{2}\bar{P}\left(  p^{0}-\frac{\bar{P}}{\sqrt{2}}\left(  x-\frac{1}
{2}\right)  \right)  =\omega+m_{\pi}^{2}\left(  \frac{1}{2}-x\right)  .
\end{equation}
The last term has been introduced in order to obtain symmetric expressions. In
terms of $\omega$ we have
\begin{align}
p^{2}  &  =\omega\left(  x-\frac{1}{2}\right)  -m_{\pi}^{2}\left(  \frac{1}
{2}-x\right)  ^{2}-\vec{p}_{\bot}^{2}~~,\nonumber\\
\left(  p-\frac{1}{2}P\right)  ^{2}  &  =-\omega\left(  1-x\right)  +m_{\pi
}^{2}x\left(  1-x\right)  -\vec{p}_{\bot}^{2}~~,\\
\left(  p+\frac{1}{2}P\right)  ^{2}  &  =\omega x+m_{\pi}^{2}\left(
1-x\right)  x-\vec{p}_{\bot}^{2}~~.\nonumber
\end{align}
Next we perform the Wick rotation in terms of the $\omega$ variable and
thereafter its integration.

The calculation is performed in Euclidean space. To bring back these results
to Minkowsky space is, in general, a non trivial matter, as will be discussed
shortly. To check the consistency our scheme we will rely on the verification
of four fundamental properties of the PDFs which arise form conservation laws,
i.e. momentum conservation, gauge symmetry, isospin symmetry and the fixed
number of valence particles in our state. These properties are: (i) the parton
distribution must be defined between $0<x<1;$ (ii) the first moment of the
parton distribution is 1; (iii) the second moment of the parton distribution
is 1/2; (iv) the parton distribution must be symmetric $q\left(  x\right)
=q\left(  1-x\right)  .$

The way the first property is realized, in those cases where an exact solution
is possible \cite{NogueraTheusslVento04}, is associated to the Wick rotation.
Let us assume that the masses are momentum independent. The denominators
present in the definition of the parton distribution, Eqs. (\ref{02.04}) and
(\ref{02.05}), can be written as
\begin{subequations}
\label{02.23}%
\begin{align}
\left(  p-\frac{1}{2}P\right)  ^{2}-m^{2}+i\varepsilon &  =-\left(
1-x\right)  \left[  \omega-m_{\pi}^{2}x+\frac{\vec{p}_{\bot}^{2}+m^{2}%
}{\left(  1-x\right)  }-i\frac{\varepsilon}{\left(  1-x\right)  }\right]
\label{02.23a}\\
\left(  p+\frac{1}{2}P\right)  ^{2}-m^{2}+i\varepsilon &  =x\left[
\omega+m_{\pi}^{2}\left(  1-x\right)  -\frac{\vec{p}_{\bot}^{2}+m^{2}}%
{x}+i\frac{\varepsilon}{x}\right]  \label{02.23b}%
\end{align}
and we observe that the integration over $\omega$ is different from zero only
if $0<x<1.$ We can perform the Wick rotation in the region $0<x<1,$ where this
rotation is well defined according to the pole positions given by Eqs.
(\ref{02.23}). For the values of $x$ in the regions $x<0$ and $x>1$ the Wick
rotation is not allowed.

The models used here are too complex to allow for a similar proof of the first
property, but we have been able to show numerically that the parton
distribution vanishes outside the interval. Moreover, the last three
properties come out of the calculation and represent our consistency check.

In many models these four properties are not satisfied. In that case authors
proceed in the following way: (i) they impose that the PDF is only valid for
$0<x<1$; (ii) they force the normalization condition of the PDF by changing
the normalization of the wave function or BS amplitude; however in this
process, the form factor of the pion will loose its normalization; (iii) some
authors claim that not satisfying the third and fourth properties is
consequence of the lack of \textquotedblleft gluon component" in the meson in
those models.

\section{Experimental analysis of PDFs.}

In order to get numerical estimates for the PDFs we need to introduce specific
models which provide us with expressions for $G_{0}\left(  p\right)  $ and
$G_{p}\left(  p\right)  $ or alternatively for $m\left(  p\right)  $ and
$Z\left(  p\right)  .$ There are many papers dedicated to the study of the
quark mass term in the quark propagator \cite{DyakonovPetrov86, Bowman02,
Bowman03, BurdenRobertsWilliams92, Roberts96, MunczekNemirovsky83,
HawesRobertsWilliams94, BenderDetmoldThomasRoberts02, RuizArriolaBroniowski03}
. We consider two different scenarios based on different philosophies but
which give quite coherent results. From now on the analysis is carried out in
Euclidean space a feature which we indicate by the sublabel $E$ in the
corresponding momenta.

The first scenario, called hereafter S1, is based on the work of Dyakonov and
Petrov \cite{DyakonovPetrov86}, which provides us with the momentum dependence
of the quark mass term coming from a instanton model. They assume $Z\left(
p_{E}\right)  =1$ and work in the chiral limit ($m_{0}=0$). Their results are
well described by the expression
\end{subequations}
\begin{equation}
m\left(  p_{E}\right)  =m_{0}+\alpha_{m}\left(  \frac{\Lambda_{m}^{2}}
{\Lambda_{m}^{2}+p_{E}^{2}}\right)  ^{3/2} \label{03.22DyakonovPetrov0}%
\end{equation}
with $\Lambda_{m}=0.767 \operatorname{GeV} $ and $\alpha_{m}=0.343
\operatorname{GeV} .$

The second scenario, S2, corresponds to an alternative mass function obtained
from lattice calculations by Bowman et al. \cite{Bowman02, Bowman03},
\begin{equation}
m\left(  p_{E}\right)  =m_{0}+\alpha_{m}\frac{\Lambda_{m}^{3}}{\Lambda_{m}
^{3}+\left(  p_{E}^{2}\right)  ^{1.5}} \label{03.23mBowman}%
\end{equation}
with $\Lambda_{m}=0.719 \operatorname{GeV} $ and $\alpha_{m}=0.302
\operatorname{GeV} .$ In their lattice analysis the authors also look for the
wave function renormalization constant. Their values are reasonably reproduced
by:
\begin{equation}
Z\left(  p_{E}\right)  =1+\alpha_{z}\left(  \frac{\Lambda_{z}^{2}}{\Lambda
_{z}^{2}+p_{E}^{2}}\right)  ^{5/2} \label{03.24ZBowman}%
\end{equation}
with $\alpha_{z}=-0.5$ and $\Lambda_{z}=1.183 \operatorname{GeV} .$

In reference \cite{Noguera05} different properties of these two scenarios have
been studied. In particular, in table \ref{Table 1} we show the values of
$\left\langle \bar{q}q\right\rangle ^{1/3},$ pion mass and mean squared radius
obtained for these scenarios. The form factor of the pion is also analyzed in
\cite{Noguera05}, obtaining a reasonable agreement.

\begin{table}[ptb]
\centering
\begin{tabular}
[c]{|c|c|c|c|c|}\hline
$\ $Case$\ $ & $\left\langle \bar{q}q\right\rangle ^{1/3}\left(
\operatorname{MeV} \right)  $ & $m_{0}\left(  \operatorname{MeV} \right)  $ &
$m_{\pi}\left(  \operatorname{MeV} \right)  $ & $<r^{2}>( \operatorname{fm}
^{2})$\\\hline
S1 & $-303.$ & $2.3$ & $137.$ & $0.41$\\\hline
S2 & $-285.$ & $3.0$ & $139.$ & $0.36$\\\hline
Exp. & $-250\sim-300 $ & $1.5 \sim8 $ & $135. \sim140.$ & $0.44$\\\hline
\end{tabular}
\caption{Results for $<qq>^{1/3}$ , $m_{0}$, the corresponding $M_{p}$ and the
rms radius squared $<r^{2}>$, for the two scenarios described in the main
text.}%
\label{Table 1}%
\end{table}

\begin{figure}[ptb]
\begin{center}
\includegraphics[height=6.6602cm,width=8.23cm]{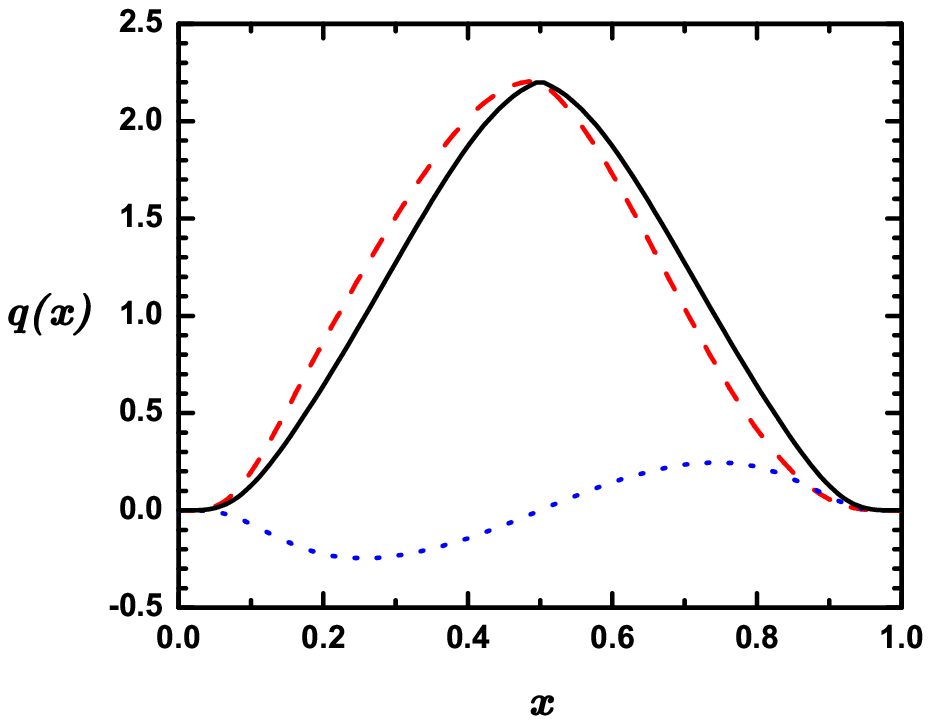} \hspace{2em}
\includegraphics[height=6.6602cm,width=8.23cm]{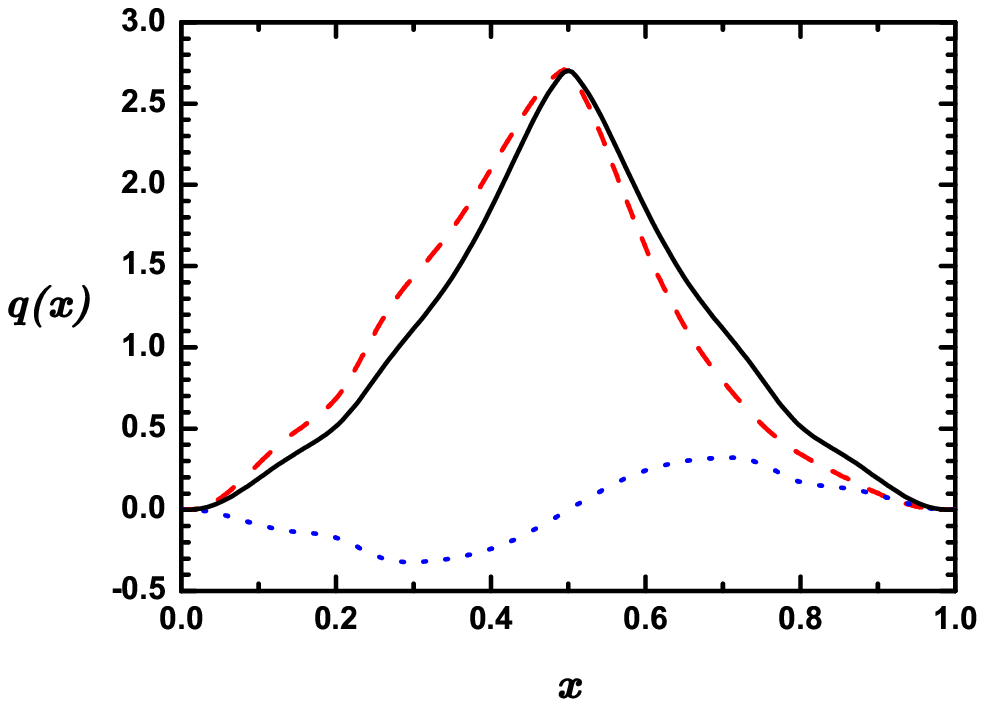}
\end{center}
\caption{PDFs for S1 (left) and S2 (right). We have drawn the contribution
from Eq. (\ref{02.04}) ($q_{u}^{\left(  1\right)  }\left(  x\right)  $, dashed
line), from Eq. (\ref{02.05}) ($\tilde{q}_{u}^{\left(  1\right)  }\left(
x\right)  $, dotted line) and the total parton distribution (full line).}%
\label{figPDCases12}%
\end{figure}

In Figs. \ref{figPDCases12} we show the parton distributions obtained for the
two defined scenarios. In both cases we have drawn the contribution from
Eq.(\ref{02.04}) ($q_{u}^{\left(  1\right)  }\left(  x\right)  $, dashed
line), from Eq. (\ref{02.05}) ($\tilde{q}_{u}^{\left(  1\right)  }\left(
x\right)  $, dotted line) and the total parton distribution (full line). We
observe that only the total parton distribution is symmetric around the point
$x=0.5$, as isospin predicts. The first moment of the parton distribution is
automatically 1, provided that our pion BS amplitude is well normalized using
Eq. (\ref{02.02}). The contribution to the first moment of the new term,
$\tilde{q}_{u}^{\left(  1\right)  }\left(  x\right)  $, is zero as a
consequence of charge conjugation symmetry \cite{Noguera05}. Regarding the
second moment of the parton distribution, we have for S1 that the contribution
from $q_{u}^{\left(  1\right)  }\left(  x\right)  $ is 0.47 and the one coming
from $\tilde{q}_{u}^{\left(  1\right)  }\left(  x\right)  $ is 0.03. In the S2
their values are are 0.46 and 0.04 respectively. We observe that our
\textquotedblleft new" contribution, $\tilde{q}_{u}^{\left(  1\right)
}\left(  x\right)  $ , restores the correct value, with a contribution at the
level of 8\%.

We have also looked to the behavior of the PDFs under variations of the
parameters present in the expression of $m\left(  p\right)  $ and $Z\left(
p\right)  $. We observe that not all values of the parameters are permitted by
our consistency requirements. For instance, if we increase $\alpha_{m}$ or
$\alpha_{Z}$ by a factor 2 we observe that the second moment of the PDF
deviates from 0.5 and the PDF is not more symmetric around the point $x=0.5.$
These results appear because the Wick rotation introduces spurious
contributions. Let us be more precise, our formalism preserves all the
symmetries up to the point in which we perform the Wick rotation. It is while
performing the latter that, if \textquotedblleft spurious" poles and/or cuts
are close to the physical ones, appreciable errors might arise. The position
of these singularities depends on the \textquotedblleft models" and the
\textquotedblleft parameter choice" in those models.

\begin{figure}[ptb]
\begin{center}
\includegraphics[
height=7.9232cm, width=9.5246cm ] {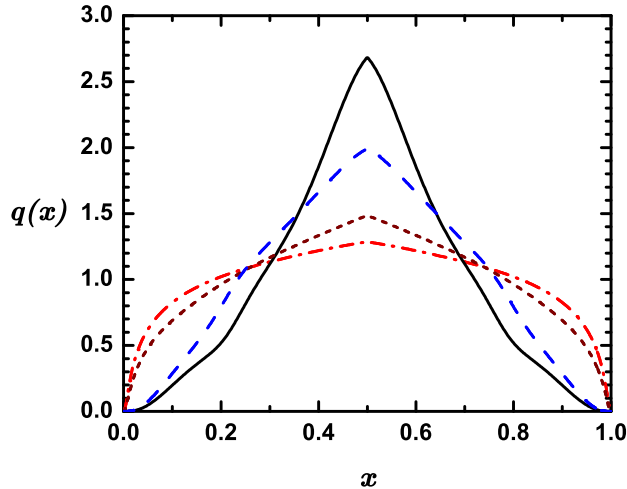}
\end{center}
\caption{Full line corresponds to S2, Eqs. (\ref{03.23mBowman} ) and
(\ref{03.24ZBowman}). The other curves corresponds to changes in the
parameters of this scenario: (i) $\alpha_{Z}$ divided by 2 and $\Lambda_{Z}$
multiplied by 2 (dashed line); (ii) $\alpha_{m}$ divided by 2 and $\Lambda
_{m}$ multiplied by 2 (dotted line); (iii) both $\alpha_{Z}$ and $\alpha_{m}$
divided by 2 and both $\Lambda_{m}$ and $\Lambda_{Z} $ multiplied by 2
(dashed-dotted line).}%
\label{FigPDCase2Comp1}%
\end{figure}

Let us now change our parameters in the opposite direction in order to compare
our approach with the NJL model. We multiply $\Lambda_{m}$ and $\Lambda_{Z}$
by 2, and at the same time divide $\alpha_{m}$ and $\alpha_{Z}$ by 2 to insure
that the value of the physical observables, like the quark condensate, do not
change too much. This change makes the currents in Eqs. (\ref{02.08Currents})
more point like, leading to a model whose results, in the limit $\alpha
_{Z}\rightarrow0,$ $\Lambda_{m}\rightarrow\infty$ resemble those of the NJL
model. In Fig. \ref{FigPDCase2Comp1} we show the pion PDF in S2, Eqs.
(\ref{03.23mBowman}) and (\ref{03.24ZBowman}), with the following changes: (i)
$\Lambda_{Z}$ multiplied by 2 and $\alpha_{Z}$ divided by 2 (dashed line);
(ii) $\Lambda_{m}$ multiplied by 2 and $\alpha_{m}$ divided by 2 (dotted
line); (iii) both $\Lambda m$ and $\Lambda_{Z}$ multiplied by 2 and both
$\alpha_{Z}$ and $\alpha_{m}$ divided by 2 (dashed-dotted line). We observe
the similarity, especially of the last case, to the result of the NJL model in
the chiral limit, $q\left(  x\right)  =1.$

Some doubts maybe cast in our calculation having to do with the relation
between the calculation in Minkowsky and Euclidean spaces, as a consequence of
the singularity structure discussed in section II. In section IV we resolve
this problem by calculating the moments of the parton distribution functions
in three different ways and proving that for the scenarios described above the
three methods fully agree.

By explicit construction, our quarks are constituent quarks. The experimental
parton distributions however, unveil the structure of the pion in terms of
fundamental partons. How to relate the constituent quarks and the fundamental
partons is an old problem analyzed by Altarelli, Cabibbo, Maiani and Petronzio
\cite{AltarelliCabibboMaianiPetronzio74}, and applied more recently to the
pion \cite{AltarelliPetrarcaRapuano96} and to the nucleon
\cite{Scopetta1,Scopetta2}. Let us recall the main features of this development.

As shown in ref.\cite{AltarelliCabibboMaianiPetronzio74}, the constituent
quarks are themselves composite objects whose structure is described by a set
of functions $\phi_{q_{0}q}\left(  x\right)  $ that specify the number of
point-like partons of type $q$ which are present in the constituent of type
$q_{0}$, with fraction $x$ of its total momentum. We will hereafter call these
functions, generically, the structure functions of the constituent quark. The
functions describing the nucleon parton distributions are expressed in terms
of these new functions\ $\phi_{q_{0} q}\left(  x\right)  $ and of the
calculated constituent density distributions $(q_{0}=u_{0},d_{0})$ as,
\begin{equation}
q\left(  x,Q^{2}\right)  =\sum_{q_{0}}\int_{x}^{1}\frac{dz}{z}\ q_{0}\left(
z,Q^{2}\right)  \ \phi_{q_{0}q}\left(  \frac{x}{z},Q^{2}\right)
\end{equation}
where $q$ labels the various partons, i.e., valence quarks and antiquarks
$(u_{v},\bar{d}_{v})$, sea quarks $(u_{s},d_{s},s)$, sea antiquarks $\left(
\bar{u}_{s},\bar{d}_{s},\bar{s}\right)  $ and gluons $g.$

The different types and functional forms of the structure functions of the
constituent quarks are derived from three very natural assumptions
\cite{AltarelliCabibboMaianiPetronzio74}: (i) The point-like partons are QCD
degrees of freedom, i.e. quarks, antiquarks and gluons; (ii) Regge behavior
for $x\rightarrow0$ and duality ideas; (iii) invariance under charge
conjugation and isospin.

These considerations define in the case of the valence quarks the following
structure function
\begin{equation}
\phi_{q_{0} q_{v}}\left(  x,Q^{2}\right)  =\frac{\Gamma\left(  A+\frac{1}
{2}\right)  }{\Gamma\left(  \frac{1}{2}\right)  \Gamma\left(  A\right)  }
\frac{\left(  1-x\right)  ^{A-1}}{\sqrt{x}}\ .
\end{equation}
For the sea quarks the corresponding structure function becomes,
\begin{equation}
\phi_{q_{0} q_{s}}\left(  x,Q^{2}\right)  =\frac{C}{x}\left(  1-x\right)
^{D-1}\ ,
\end{equation}
and, in the case of the gluons, it is taken
\begin{equation}
\phi_{q_{0} g}\left(  x,Q^{2}\right)  =\frac{G}{x}\left(  1-x\right)
^{B-1}\ .
\end{equation}
We take the same parameters as in ref.\cite{Scopetta1} for the proton case
$A=0.435,$\ $B=0.378,$ \ $C=0.05,$\ $D=2.778,$ and $G=0.135$, since these
functions should in principle be independent of the hadron under scrutiny.

Once the convolution defined by Altarelli et al. is performed, we proceed to
look for the evolution of the parton distribution under the Renormalization
Group. The evolution is carried to next to leading order. We find the hadronic
scale, i.e. the scale at which the constituent quark structure is defined, at
$Q_{0}^{2}=1 \operatorname{GeV} ^{2}$. This scale is determined by imposing
that the fraction of the total momentum carried by the valence quark at
$Q^{2}=4 \operatorname{GeV} ^{2}$ is 0.46 as determined from experiment
\cite{SuttonMartinRobertsStirling92}.

\begin{figure}[ptb]
\begin{center}
\includegraphics[
height=7.9232cm, width=9.1402cm ] {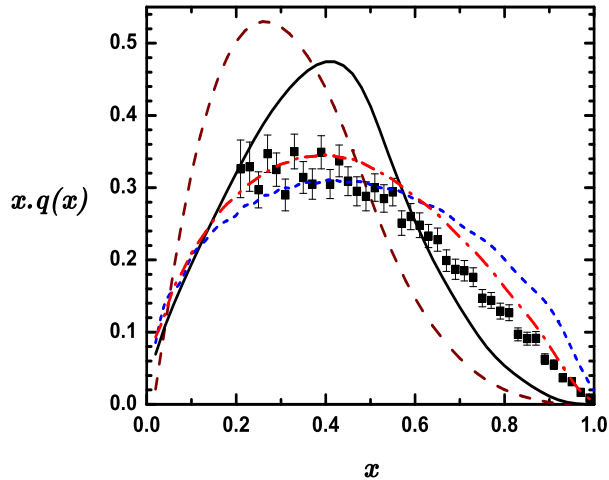}
\end{center}
\caption{Full line corresponds to the evolved PDF for S2 including the ACMP
procedure. The dashed line corresponds to the evolution of the PDF of the same
scenario without the ACMP convolution. The dashed-dotted and small-dashed
lines are for the evolved PDF for S2 with $\alpha_{Z}$ and $\alpha_{m}$
divided by 2 and $\Lambda_{m}$ and $\Lambda_{Z} $ multiplied by 2 and for the
NJL model respectively, in both cases the ACMP procedure is implemented.}%
\label{FigPDevolve1}%
\end{figure}

In Fig. \ref{FigPDevolve1} we show the evolved PDF for S2 (full line). S1,
discussed at the beginning of the section, gives similar results. The dashed
line corresponds to the evolution of the PDF of S2 without the ACMP
convolution. We observe that the results are reasonable since not a single
parameter was fixed in the fit, although too small in the $x\sim1$ region. The
ACMP procedure pushes the PDF to higher values of $x$ and therefore produces a
significant corrections in the right direction.

We show in Fig. \ref{FigPDevolve1} also the results obtained with S2 for
$\Lambda_{m}$ and $\Lambda_{Z}$ multiplied by 2 and $\alpha_{Z}$ and
$\alpha_{m}$ divided by 2 (dashed-dotted line). We observe that the
experimental results are well reproduced in this case and our curve is very
well reproduced by the expression
\begin{equation}
xq\left(  x\right)  =\frac{\Gamma\left(  1+\alpha+\beta\right)  }{
\Gamma\left(  \alpha\right)  \Gamma\left(  1+\beta\right)  } x^{\alpha}
\left(  1-x\right)  ^{\beta}\qquad\text{with }\alpha=0.62,\beta=1.04
\end{equation}
which is quite close to the bests fits in \cite{SuttonMartinRobertsStirling92}
. The fact that the PDF in terms of the constituent quarks is larger in the
regions $x\sim0$ and $x\sim1$ leads, after applying the ACMP procedure, to a
result which is reasonable in the large $x$ region. The NJL model, when the
ACMP procedure is applied to it, leads to a PDF which too large in the
$x\sim1$ region, although quite reasonable given the simplicity of the model
and the fact that no single parameter was fitted in the process (small dashed
line in Fig. \ref{FigPDevolve1}).

\section{Moments of the constituent parton distribution functions.}

In section II we have calculated the parton distribution $q\left(  x\right)
.$ We have assumed that the parton distribution is defined between $0<x<1$ and
have used as a consistency test of this assumption that the first moment of
the parton distribution is $1$ and the second moment is $1/2$. From Eq.
(\ref{02.04}), (\ref{02.05}) and (\ref{02.06})it can be analitically proved
that
\begin{equation}%
\begin{array}
[c]{c}%
\int dx~q\left(  x\right)  =F\left(  0\right)  \;\;,\\
\int dx~x~q\left(  x\right)  =\frac{1}{2}F\left(  0\right)  \;\;,
\end{array}
\label{IV.00}%
\end{equation}
when the integration is not restricted to the region $0 \le x \le1$
\footnote{In \cite{Noguera05} is was proved that the pion form factor appears
correctly normalized, $F\left(  0\right)  =1$, in the formalism.}. The
consistency test proposed in section II reflects that the significant
contribution to the two integrals (\ref{IV.00}) comes from the physical region
$0 \le x \le1$.

The numerical value of the higher moments are not fixed by symmetry arguments.
Nevertheless we can use them as a proof of the consistency of our procedure if
we are able to calculate them in a different way. We proceed to do so.

The moments of the parton distribution are defined as:
\begin{equation}
\left\langle x\right\rangle _{s}=\int_{0}^{1}dx~x^{s-1}~q\left(  x\right)
\label{IV.01}%
\end{equation}
A first procedure to calculate the moments is to perform the integration
defined in Eq. (\ref{IV.01}) using the parton distribution, $q\left(
x\right)  ,$ obtained in section II. An alternative method is to perform the
integral over $x,$ using the delta functions present in Eqs. (\ref{02.04}),
(\ref{02.05}) and (\ref{02.06}), before performing the integration over the
internal momentum, $p,$ in these equations. We focus our discussion in the
contribution to $\left\langle x\right\rangle _{s}$ coming from (\ref{02.04}),
because the generalization to the other two contributions is straightforward.

Once the integral over $x$ is performed, the contribution to $\left\langle
x\right\rangle _{s}$ coming from (\ref{02.04}) is%

\begin{gather}
\left\langle x_{1}\right\rangle _{s}=-\int\frac{d^{4}p}{\left(  2\pi\right)
^{4}}\text{ }\left[  \frac{n}{2}\cdot\left(  2p+P\right)  \right]
^{s-1}\,\mathbb{T}\mathrm{r}\left[  i\,S\left(  p-\frac{1}{2}P\right)
\bar{\Gamma}^{M}\left(  p,P\right)  i\,S\left(  p+\frac{1}{2}P\right)  \right.
\nonumber\\
\left.  \Gamma_{\mu}\left(  p+\frac{1}{2}P,p+\frac{1}{2}P\right)  n^{\mu
}\,i\,S\left(  p+\frac{1}{2}P\right)  \Gamma^{M}\left(  p,P\right)  \right]
\label{IV.02}%
\end{gather}
with the momentum integration restricted to those values that guaranty the condition%

\begin{equation}
0<\frac{n}{2}\cdot\left(  2p+P\right)  <1\;\;. \label{IV.02a}%
\end{equation}
The difference between the first and the second procedures is just the order
of integration of the variables. Any significant problem with singularities
must appear in comparing the result of these two methods.

For calculating the integral present in Eq. (\ref{IV.02}) we need a reference
frame. The most adequate one is the infinite momentum frame, in which we have
$n^{\mu}=\left(  1,0,0,-1\right)  /\sqrt{2}\bar{P}$ with $\bar{P} $ going to
infinity. The pion momentum is expressed as%

\begin{equation}
P^{\mu}=\frac{1}{\sqrt{2}}\left(  \bar{P}+\frac{m_{\pi}^{2}}{2\bar{P}
},0,0,\bar{P}-\frac{m_{\pi}^{2}}{2\bar{P}}\right)  . \label{IV.03}%
\end{equation}
In this reference frame the dominant terms appearing in the denominators of
the propagators are proportional to $\bar{P},$ $[~\left(  p\pm P/2\right)
^{2} $ $\propto$ $\pm\frac{1}{\sqrt{2}}\left(  p^{0}-p^{3}\right)  \bar{P}~],
$\ which lead to an integrand vanishing as $\bar{P}^{-2}$ for large values of
$\bar{P},$ except in a small region of $p^{0}$ around $p^{3}.$ In order to
evaluate the contribution of this region it is adequate to make the following
change of variable%

\begin{equation}
p^{0}=p^{3}+\frac{\bar{m}}{\bar{P}}p^{\prime0} \label{IV.04}%
\end{equation}
where $\bar{m}$ is any mass in the problem, for instance $\bar{m}=m\left(
0\right)  .$

Using (\ref{IV.04}) in eq. (\ref{IV.02a}) we obtain that the integration over
$p^{3}$ must be performed between $-\bar{P}/(2\sqrt{2})$ and $\bar{P}
/(2\sqrt{2})$ and similarly for $p^{0}$. Then, the contribution coming from
(\ref{IV.02}) is%

\begin{gather}
\left\langle x_{1}\right\rangle _{s}=-\lim_{\bar{P}\rightarrow\infty}
\int_{-\bar{P}/(2\sqrt{2})}^{\bar{P}/(2\sqrt{2})}\frac{\bar{m}}{\bar{P}}
\frac{dp^{\prime0}}{2\pi}\int_{-\bar{P}/(2\sqrt{2})}^{\bar{P}/(2\sqrt{2}
)}\frac{dp^{3}}{2\pi}\int\frac{d^{2}p_{\bot}}{\left(  2\pi\right)  ^{4}}\text{
}\left[  \sqrt{2}\frac{p^{3}}{\bar{P}}+\frac{1}{2}\right]  ^{m-1}\,\nonumber\\
\mathbb{T}\mathrm{r}\left[  i\,S\left(  p-\frac{1}{2}P\right)  \bar{\Gamma
}^{M}\left(  p,P\right)  i\,S\left(  p+\frac{1}{2}P\right)  \right.  \left.
\Gamma_{\mu}\left(  p+\frac{1}{2}P,p+\frac{1}{2}P\right)  n^{\mu}\,i\,S\left(
p+\frac{1}{2}P\right)  \Gamma^{M}\left(  p,P\right)  \right]  \label{IV.05}%
\end{gather}
It is a this stage that we assume that the Wick rotation on the variable
$p^{\prime0}$ can be performed without problems. The most dangerous term,
$\left[  n\cdot\left(  2p+P\right)  /2\right]  ^{s-1},$ is under control for
any value of $s$. It is not difficult to show that this integral has a well
defined value as $\bar{P}\rightarrow\infty$ and that this value is independent
of $\bar{m}$ in Eq.(\ref{IV.04}).

We have applied the same procedure to the other two contributions to
$\left\langle x\right\rangle _{s},$ coming form Eqs. (\ref{02.05}) and
(\ref{02.06}) and we have calculated the moments of the parton distribution in
the two scenarios, S1 and S2, defined in section III. We have also calculated
them directly from Eq. (\ref{IV.01}) with the parton distributions obtained in
section III. The numerical results coincide in both scenarios for all the moments.

Summarizing, in here we have calculated the moments of the parton distribution
functions for the two scenarios defined in section III in two different ways.
Firstly, from Eq. (\ref{IV.01}) using the calculated parton distribution
obtained from the method explained in section II. In this first method we do
not need to define a particular frame for the calculation. In the second
method, we have calculated the moments directly in Euclidean space and in the
infinite momentum frame. This second procedure has been justified from an
analysis in Minkowsky space, by performing the Wick rotation in Eq.
(\ref{IV.05}) and the equivalent equations for the other two contributions
coming from Eqs. (\ref{02.05}) and (\ref{02.06}). The methods are based on
significantly different calculations and therefore, they provide a test of
consistency. Those models which violate the check of consistency defined at
the end of section II, will also give different numerical results for the
moments of the parton distribution functions in the two methods.

\section{Conclusions.}

We have applied a procedure for calculating the PDFs in a framework, based on
the Dyson-Schwinger equations, first introduced in ref. \cite{Noguera05}. The
motivation behind our procedure is the preservation of all the conservation
laws associated with the fundamental symmetries (Poincar\'{e} covariance,
gauge invariance, isospin symmetry and number of valence constituents) which
implies that before evolution the PDF of the pion built with only two valence
(anti)quarks has first moment equal to 1, second moment equal to 0.5, is
symmetric around the point $x=0.5$ and is defined in $0<x<1.$ Our procedure,
determined by the appropriate implementation of local gauge invariance, is
defined in terms of an operator, which incorporates additional contributions
besides the conventional triangle diagram \cite{Noguera05}. The method of
calculation follows the method used in \cite{NogueraTheusslVento04} for the
calculation of the GPDs in the NJL model and can be easily applied to any
model with similar ingredients.

Our approach preserves all the symmetries up to the moment in which the Wick
rotation is performed. The particular choice for $m\left(  p\right)  $ and
$Z\left(  p\right)  $ becomes crucial at this moment. An arbitrary choice of
$m\left(  p\right)  $ or $Z\left(  p\right)  $ can have \textquotedblleft
spurious" singularities which might lead to appreciable symmetry breaking
effects. Our expressions for these functions, obtained from QCD studies, are
consistent, within our numerical precision, with all the symmetries. Note the
parton distribution vanishes at the support endpoints, $x=0$ and $x=1$, a
feature which occurs in the instanton model \cite{PolyakovWeiss99}, but not in
the NJL model. The new terms in the operator give no contribution to the first
moment of the PDF, but a significant contribution ($\sim6-8\%$) to the second
moment and restore the symmetry of the PDF around the point $x=0.5 $. We can
approach the results of the NJL model by a change in the parameters.

In order to check furthermore our procedure we have calculated the moments of
the parton distribution in two different ways. The consistency of these
procedures provides a strong test for our formalism to determine the parton distributions.

In order to describe the experimental data we have realized that the ACMP
procedure is necessary. Nevertheless, even incorporating the ACMP
convolutions, the results with the given parameterization are not spectacular,
they fall very low for large $x$. On the other hand, the NJL model, also with
ACMP convolutions, behaves oppositely, its PDFs is too large in that region.
Since we can approach the NJL model in our scheme by increasing $\Lambda_{m}$
we do so, changing simultaneously $\alpha_{m}$ to keep the low energy
properties as close as possible to the experimental ones. The improvement is
immediate although we have done no fine tuning. Nevertheless, we do not make a
big statement about this improvement, since there might be other contributions
to the PDFs arising from improvements to our lagrangian, which are lacking in
our formulation.

Fundamental studies of QCD provide results which lead to effective nonlocal
interactions which are difficult to connect with the data. The analysis
carried out here shows that the formalism developed in ref.\cite{Noguera05} is
well suited for this purpose. The formalism can be easily generalized to any
model, is based on the Dyson-Schwinger equations, and therefore contains all
the non perturbative input associated with the bound state structure of
hadrons, and is suited to preserve all the wishful symmetries, as long as we
are able to control the passage from Euclidean to Minkowski space-times.
Besides these general conclusions, we have also shown, in the calculation of
the PDFs, that the models thus far developed to describe the structure of
hadrons at low $Q^{2}$ (scenarios S1 and S2) require from a convolution
formalism of the ACMP type to approach the data. Moreover, surprisingly
enough, without any change in the parameters of these models, the results,
once ACMP is incorporated, are qualitatively correct. Finally, with little
changes in the parameterization we have also shown that one can easily adjust
both low energy and high energy observables within the same model lagrangian.
We have not pretended to do so precisely, since our model is far too crude to
be able to explain the data, but our calculations shows that this enterprise
is feasible.

\begin{acknowledgments}
We would like to thank Dr. Sergio Scopetta for his tuition in the use of the
evolution code. This work was supported by the sixth framework programme of
the European Commission under contract 506078 (I3 Hadron Physics), MEC (Spain)
under contracts BFM2001-3563-C02-01 and FPA2004-05616-C02-01, and Generalitat
Valenciana under contract GRUPOS03/094.
\end{acknowledgments}

\appendix{}

\section{Description of the quark propagator and pion amplitude.}

\label{AppendixA}

In ref \cite{Noguera05} it is studied the model described by the lagrangian of
Eq. (\ref{02.07Lagrangiano}). We obtain from the Dyson equation the momentum
dependence of the mass and wave function renormalization of the propagator of
Eq. (\ref{02.11Propagador}):%

\begin{align}
m\left(  p\right)   &  =\frac{m_{0}+\alpha_{0}G_{0}\left(  p\right)
}{1-\alpha_{p}G_{p}\left(  p\right)  }\\
Z\left(  p\right)   &  =\frac{1}{1-\alpha_{p}G_{p}\left(  p\right)  }%
\end{align}
The constants $\alpha_{0}$ and $\alpha_{p}$ are directly related to couplings
constants $g_{0}$ and $g_{p}:$
\begin{align}
\alpha_{0}  &  =i\,8\,N_{c}N_{f}g_{0}\int\frac{d^{4}p}{\left(  2\pi\right)
^{4}}G_{0}\left(  p\right)  \frac{Z\left(  p\right)  m\left(  p\right)
}{p^{2}-m^{2}\left(  p\right)  +i\epsilon}\\
\alpha_{p}  &  =i\,8\,N_{c}N_{f}g_{p}\int\frac{d^{4}p}{\left(  2\pi\right)
^{4}}G_{p}\left(  p\right)  \frac{p^{2}Z\left(  p\right)  }{p^{2}-m^{2}\left(
p\right)  +i\epsilon}%
\end{align}

The BS equation is solved leading to:
\begin{equation}
\Gamma^{M=\pi^{i}}\left(  p,P\right)  =\,i\,\gamma_{5}\,\tau^{i}\,g_{\pi
qq}G_{0}\left(  p\right)
\end{equation}
Defining the pseudo-scalar polarization
\begin{equation}
\Pi_{ps}\left(  P^{2}\right)  =-~i\,4\,N_{c}\,N_{f}\int\frac{d^{4}p}{\left(
2\pi\right)  ^{4}}G_{0}^{2}\left(  p\right)  \frac{Z\left(  p+\frac{1}
{2}P\right)  \,Z\left(  p-\frac{1}{2}P\right)  \left[  \frac{1}{4}P^{2}
-p^{2}+m\left(  p+\frac{1}{2}P\right)  m\left(  p-\frac{1}{2}P\right)
\right]  }{\left[  \left(  p+\frac{1}{2}P\right)  ^{2}-m^{2}\left(  p+\frac
{1}{2}P\right)  +i\epsilon\right]  \left[  \left(  p-\frac{1}{2}P\right)
^{2}-m^{2}\left(  p-\frac{1}{2}P\right)  +i\epsilon\right]  }~~,
\end{equation}
the pion mass is obtained from the relation:
\begin{equation}
1=2\,g_{0}\,\Pi_{ps}\left(  P^{2}=m_{\pi}^{2}\right)
\end{equation}
and the normalization constant $g_{\pi qq}$ is given by
\begin{equation}
\frac{1}{g_{\pi qq}^{2}}=-\left(  \frac{\partial\Pi_{ps}}{\partial P^{2}
}\right)  _{P^{2}=m_{\pi}^{2}}~.
\end{equation}


\begin{thebibliography}{99}                                                                                               %


\bibitem {MullerRobaschikGeyerDittesHorejsiFortsch94-98}D. M\"{u}ller, D.
Robaschick, B. Geyer, F.M. Dittes and J. Horejsi,Fortschr. Phys. \textbf{42},
101 (1994).

\bibitem {Radyushkin97}A.V. Radyushkin, Phys. Lett \textbf{B380}, 417 (1996),
Phys. Lett \textbf{B385}, 333 (1996).

\bibitem {Ji97}X.D. Ji, Phys. Rev. Lett. \textbf{78}, 610 (1997), Phys. Rev.
\textbf{D55}, 7114 (1997).

\bibitem {Diehl03}M. Diehl, Phys. Rep. \textbf{388}, 41 (2003).

\bibitem {Noguera05}S. Noguera, hep-ph/0502171

\bibitem {BadierNA3}J. Badier et al. NA3 Collaboration, Z. Phys. \textbf{C18},
281 (1983)

\bibitem {BetevNA10}B. Betev et al. NA10 Collaboration Z. Phys. \textbf{C28},
15 (1985)

\bibitem {Conway89}J. S. Conway et al Phys Rev \textbf{D39}, 92 (1989)

\bibitem {Aurencheetall89}P. Aurenche et al. Phys. Lett. \textbf{B233}, 517 (1989)

\bibitem {SuttonMartinRobertsStirling92}P. J. Sutton, A. D. Martin, R. G.
Roberts, W. J. Stirling, Phys. Rev. \textbf{D45}, 2349 (1992)

\bibitem {BestSchafer97}C. Best et al.Phys. Rev. \textbf{D56}, 2743 (1997).

\bibitem {DetmoldMelnitchoukThomas03}W. Detmold, W. Melnichouk, A. W. Thomas,
Phys. Rev. D68, 034025 (2003)

\bibitem {DavidsonRuizArriola}R. M. Davidson and E. Ruiz Arriola, Phys. Lett.
B348, 163 (1995); Acta Phys. Pol. \textbf{B33}, 1791 (2002).

\bibitem {Lansberg}F. Bissey, J.R. Cudell, J. Cugnon, J.P. Lansberg, P.
Stassart, Phys.Lett. \textbf{B587}, 189 (2004); F. Bissey, J.R. Cudell, J.
Cugnon, M. Jaminon, J.P. Lansberg, P. Stassart, Phys.Lett. \textbf{B547} 210 (2002).

\bibitem {RuizArriolaBroniowski02}E. Ruiz Arriola and W. Broniowski, Phys.
Rev. \textbf{D66}, 094016 (2002).

\bibitem {NogueraTheusslVento04}S. Noguera, L. Theu\ss l and V. Vento
Eur.Phys.J. \textbf{A20}, 483 (2004).

\bibitem {Nambu}Y. Nambu and G. Jona-Lasinio, Phys. Rev. \textbf{124}, 246 (1961).

\bibitem {RuizArriolaBroniowski03}E. Ruiz Arriola and W. Broniowski, Phys.
Rev. D \textbf{67}, 074021 (2003).

\bibitem {PolyakovWeiss99}M. V. Polyakov and C. Weiss, Phys. Rev. D60, 114017 (1999)

\bibitem {Polyakov99}M. V. Polyakov Nucl. Phys. B555, 231 (1999)

\bibitem {AlbergHenley05}M. Alberg and E. M. Henley, Phys. Lett.
\textbf{B611}, 111 (2005)

\bibitem {AnikinBinosiMedranoNogueraVento02}I.V. Anikin, D. Binosi, R.
Medrano, S. Noguera, V. Vento Eur.Phys.J.\textbf{A14}, 95 (2002).

\bibitem {TiburziMiller}B. C. Tiburzi and G. Miller, Phys. Rev. \textbf{C64},
065204 (2001); Phys. Rev. \textbf{D65}, 074009 (2002); Phys. Rev.
\textbf{D67}, 113004 (2003). B. C. Tiburzi Ph.\ D. Thesis nucl-th/0407005

\bibitem {Simula}S. Simula hep-ph/0406074

\bibitem {HechtRobertsSchmidt01}M. B. Hecht, C. D. Roberts and S. M. Schmidt,
Phys. Rev. \textbf{C63}, 025213 (2001).

\bibitem {Traini}M. Traini, V. Vento, A. Mair, A. Zambarda, Nucl. Phys. A614,
472 (1997).

\bibitem {AltarelliCabibboMaianiPetronzio74}G. Altarelli, N. Cabibbo, L.
Maiani, R. Petronzio, Nucl. Phys. \textbf{B69}, 531 (1974).

\bibitem {Scopetta1}S. Scopetta, V. Vento, M. Traini, Phys. Lett.
\textbf{B421}, 64 (1998); Phys. Lett. \textbf{B442}, 28 (1998).

\bibitem {Scopetta2}S. Scopetta, V. Vento, Phys. Rev. \textbf{D69}, 094004
(2004); Phys.Rev. \textbf{D71}, 014014 (2005).

\bibitem {AltarelliPetrarcaRapuano96}G. Altarelli, S. Petrarca, F. Rapuano,
Phys. Lett. \textbf{B373}, 200 (1996).

\bibitem {NogueraScopettaVento}S. Noguera, S. Scopetta, V. Vento, Phys. Rev.
\textbf{D70}, 094018 (2004).

\bibitem {BallChiu80}J.S. Ball and T.W. Chiu, Phys. Rev. \textbf{D22}, 2542 (1980).

\bibitem {CurtisPennington90}D.C. Curtis and M.R. Pennington, Phys. Rev.
\textbf{D42}, 4165 (1990).

\bibitem {Roberts96}C. D. Roberts, Nucl. Phys. \textbf{A605}, 475\ (1996).

\bibitem {BenderDetmoldThomasRoberts02}A. Bender, W. Detmold, A. W. Thomas ,
C.D. Roberts, Phys.Rev. \textbf{C65}, 065203, (2002).

\bibitem {DyakonovPetrov86}D. I. Dyakonov and V. Yu. Petrov, Nucl. Phys.
\textbf{B272}, 457 (1986).

\bibitem {Bowman02}P. O. Bowman, U. M. Heller and A. G. Williams, Phys. Rev.
\textbf{D66}, 014505 (2002) .

\bibitem {Bowman03}P. O. Bowman, U. M. Heller, D. B. Leinweber and A. G.
Williams, Nucl.Phys.Proc.Suppl.\textbf{119}, 323 (2003).

\bibitem {BurdenRobertsWilliams92}C. J. Burden, C. D. Roberts and A. G.
Williams, Phys. Lett. \textbf{B285}, 347 (1992).

\bibitem {MunczekNemirovsky83}H. J. Munczek and A. M. Nemirovsky, Phys. Rev.
D\textbf{\ 28}, 181 (1983).

\bibitem {HawesRobertsWilliams94}F. T. Hawes, C. D. Roberts and A. G.
Williams, Phys. Rev. D\textbf{\ 49}, 4683 (1994).
\end{thebibliography}
\end{document}